\documentclass[twocolumn,prl,aps]{revtex4-1}
\usepackage{graphicx}
\usepackage{amsmath}
\usepackage{amssymb}
\usepackage{natbib}

\def\ket#1{{\left| #1 \right\rangle}}

\begin{document}

%\title{An evolutionary algorithm to engineer quantum states with an application in optical quantum metrology}
\title{A search algorithm for quantum state engineering and metrology}
\author{P. A. Knott}
	\email{P.Knott@Sussex.ac.uk}
	\affiliation{Department of Physics and Astronomy, University of Sussex, Brighton BN1 9QH, United Kingdom}

%\pacs{42.50.St,42.50.Dv,03.65.Ud,03.65.Ta,06.20.Dk}

\date{\today}

%================Abstract===================%
\begin{abstract}
In this paper we present a search algorithm that finds useful optical quantum states which can be created with current technology. We apply the algorithm to the field of quantum metrology with the goal of finding states that can measure a phase shift to a high precision. Our algorithm efficiently produces a number of novel solutions: we find experimentally-ready schemes to produce states that show significant improvements over the state-of-the-art, and can measure with a precision that beats the shot noise limit by over a factor of $4$. Furthermore, these states demonstrate a robustness to moderate/high photon losses, and we present a conceptually simple measurement scheme that saturates the Cram\'er-Rao bound.

%With the fast growing interest in quantum technologies it is becoming increasingly important to engineer quantum states for specific applications. We focus on the field of optical quantum metrology: our goal is to find quantum states that can measure a phase shift to a high precision. We approach this problem by designing a search algorithm that finds optical quantum states which have the desired properties and can be created with current technology. Here our algorithm...

%With the fast growing interest in quantum technologies it is becoming increasingly important to engineer quantum states with specific properties. We design an evolutionary algorithm for this purpose that produces quantum states that can be created with current technology and can have the properties which are desired by the user. We apply this to the field of optical quantum metrology which has the goal of measuring a phase shift to a high precision. The evolutionary algorithm has found states that show significant improvements over the state-of-the-art, and can measure with a precision that improve over the shot noise limit by over a factor of $4$. Furthermore, the states we find can be made experimentally with today's technology, and are robust in the presence of moderate to high photon losses.
\end{abstract}
\maketitle

%===========Fig:================
%\begin{figure}%[h]
%\centering
%\includegraphics[scale=1]{.pdf}
%\caption{}
%\label{fig:}
%\end{figure}
%====================================

%===========Introduction==============%

\section{Introduction}

The field of optical quantum information has the potential to transform technology with a broad range of applications, including quantum computing \cite{lloyd1999quantum,kok2010introduction}, quantum cryptography \cite{hillery2000quantum} and quantum metrology \cite{demkowicz2015quantum,giovannetti2011advances}. In order for these applications to benefit from quantum-enhanced performance, non-classical states of light must be prepared, and it is an ongoing challenge to find methods to engineer quantum states with the desired properties \cite{bimbard2010quantum,ourjoumtsev2006quantum,gerrits2010generation,yukawa2013generating,bartley2012multiphoton,carolan2015universal}. In this paper we explore an alternative approach and design an algorithm to find experimentally accessible methods of engineering states with the properties we require. Here we focus on using the algorithm for quantum metrology, a field which has been instrumental in enhancing the precision of gravitational wave detectors \cite{abadie2011gravitational,aasi2013enhanced,schnabel2010quantum} and can be used to measure biological systems with minimal disturbance \cite{carlton2010fast,taylor2013biological,crespi2012measuring,whittaker2015quantum}. As with a recent (independent) algorithm that performed an automated search for new quantum experiments \cite{krenn2015automated}, our algorithm produces a range of novel, and sometimes counter-intuitive, solutions.

\begin{figure}%[h]
\centering
\includegraphics[scale=0.58]{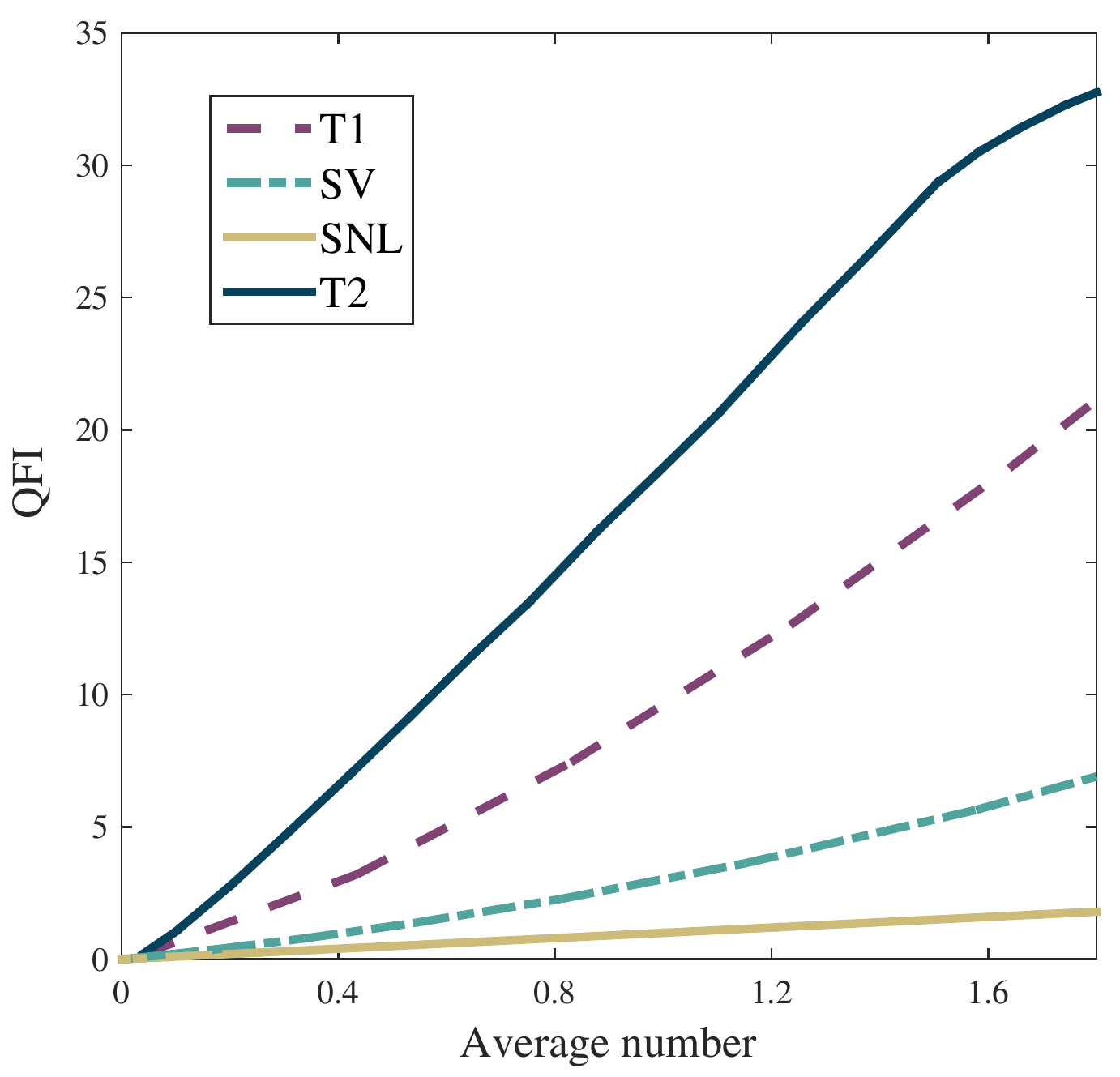}
\caption{Here we compare different states using the quantum Fisher information (QFI). The QFI is plotted against the average photon number ($\bar{n}$) for two of the quantum states found by our algorithm: $|\psi_{T1}\rangle$ and $|\psi_{T2}\rangle$ (labelled T1 and T2). A state with a large QFI can measure to a high precision, and hence $|\psi_{T1}\rangle$ and $|\psi_{T2}\rangle$ give significant improvements over the squeezed vacuum (SV) and the classical shot noise limit (SNL).}
\label{fig:QFI_B1_6N1}
\end{figure}

In this paper our aim is to find experimentally realisable states, with small photon numbers, that can measure a phase shift to a high precision. This is important for measurements on fragile systems such as spin ensembles \cite{wolfgramm2013entanglement}, biological systems \cite{carlton2010fast,taylor2013biological,crespi2012measuring,whittaker2015quantum} and atoms \cite{tey2008strong,eckert2008quantum}. There is a large literature on the theory and experiment of creating non-classical optical states for quantum metrology, and a wide range of states have been considered including NOON states \cite{lee2002quantum,afek2010high,jones2009magnetic}, squeezed states \cite{caves1981quantum,pezze2008mach,vahlbruch2010geo,mehmet2011squeezed}, and more recently squeezed cat states and squeezed entangled states \cite{knott2015practical,ourjoumtsev2007generation,huang2015optical,etesse2015experimental,lee2015quantum}. The results of our algorithm surpass all of these: we find a number of states which attain a precision that beats the classical shot noise limit (SNL) by more than a factor of 4, and can improve over the precision attainable by all other practical states (known to the author) by at least a factor of $\sqrt{6}$ (see Fig.~\ref{fig:QFI_B1_6N1}). The states we introduce can be made with today's technology, are robust to moderate/high photon losses, and can be used to measure a phase shift with a conceptually simple measurement scheme that saturates the Cram\'er-Rao bound.

%A wide range of optical states have been engineered with varying properties \cite{ourjoumtsev2006quantum,gerrits2010generation,yukawa2013generating,bartley2012multiphoton,carolan2015universal}, and there are many theoretical schemes devising new states \cite{dell2006multiphoton,park2015conditional,lee2010quantum}

%Despite this the new quantum states and techniques that have been developed only give small advantages over the original squeezed-state based scheme proposed by Carlton Caves in 1981 \cite{caves1981quantum}. Our algorithm breaks this mould by introducing states that improve over the precision attainable by all other practical states (known to the authors) by at least a factor of $\sqrt{6}$, and they beat the classical shot noise limit (SNL) by more than a factor of 4 (see Fig.~\ref{fig:QFI_B1_6N1}). Crucially our states can be made with today's technology and are robust to moderate to large photon losses.

%The states produced by our algorithm surpass all of these: we find a number of states that can be made with today's technology, which attain a precision that beats the classical shot noise limit (SNL) by more than a factor of 4, and we show substantial improvement over the precision attainable by all other practical states (known to the authors) by at least a factor of $\sqrt{6}$ (see Fig.~\ref{fig:QFI_B1_6N1}).

%===========Fig:QFI_B1_6N1================
%====================================

% What we do: state eng, toolbox, Tachikoma, metrology, T1-6 cf state-of-the-art, details and loss

%This paper is arranged as follows: we begin by introducing a practical quantum optics toolbox which consists of experimentally realised states, operators and measurements that can be used to make non-classical states. We then describe our evolutionary algorithm which combines from the toolbox to create quantum states with specific properties. We then introduce optical quantum metrology, and present a selection of states found by the evolutionary algorithm suitable for this application. We analyse the phase-measuring performance of these states as compared to the state of the art and then look in more detail at the phase-measuring scheme and we see how photon losses can affect the attainable precision.

%\section{Results}

%================== QO Toolbox ===================%

\section{A quantum state engineering algorithm}

\begin{figure}%[h]%use \begin{figure*} and \end{figure*} for full page width diagram
\centering
\includegraphics[scale=1.55]{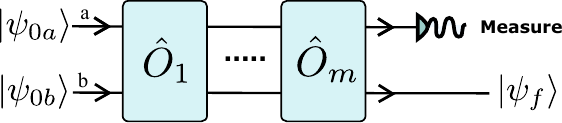}
\caption{The state engineering scheme we consider begins with two input states, $|\psi_{0a}\rangle$ and $|\psi_{0b}\rangle$, which are input into the two modes. The states then subsequently pass through a number of operators $\hat{O}_i$. To produce the final quantum state $|\psi_f\rangle$ a post-selection measurement is performed on one mode.}
\label{fig:Random_general_scheme}
\end{figure}

The state-engineering scheme we use is shown in Fig.~\ref{fig:Random_general_scheme}. Firstly, two input states, $|\psi_{0a}\rangle$ and $|\psi_{0b}\rangle$, are input into the two modes. The two modes then pass through a sequence of operators $\hat{O}_i$ where $i =1,..,m$, and the final step is to perform a post-selection measurement on one mode, producing the final state $|\psi_f\rangle$. With appropriate choices of input states, operators, and measurements this scheme is able to replicate a wide range of the quantum state engineering protocols in the literature, such as \cite{ourjoumtsev2007generation,huang2015optical,etesse2015experimental,bartley2012multiphoton,gerrits2010generation}. The main difference between the present paper and the current literature is that we don't \textit{choose} the input states, operators, and measurements, but instead allow a simulation to perform a random search for states with the desired properties. We use this technique because systematically sorting through all possible combinations of input states, operators, and measurements is not possible with current computing power - it can be shown that the number of combinations soon becomes intractable. Contrary to this, our algorithm quickly and efficiently provides new and sometimes counter-intuitive methods to create useful quantum states.

As our objective is to find \emph{practical} states for quantum metrology, we construct our state engineering protocols from elements of an experimentally-ready toolbox of quantum optics states, operators, and measurements, which is summarised in table \ref{tab:toolbox}. Here we only introduce the important details of the toolbox; more details can be found in Appendix A. Firstly, the input states we include are the squeezed vacuum $|z\rangle$, the coherent state $|\alpha\rangle$, and Fock states $|n\rangle$; the parameters $z$, $\alpha$ and $n$ are constrained by what is possible experimentally \cite{mehmet2011squeezed,muller2015coherent,claudon2010highly,morin2012high,ourjoumtsev2006quantum,huang2015thesis}.

We next introduce the operators, of which the most important is the beam splitter $\hat{U}_{T}$, where $T$ is the probability of transmission (in \%), which serves to mix and entangle the two modes. Without this the final state would just be a set of single-mode linear operators acting on an input state. The states this would create, such as displaced Fock states and squeezed coherent states, have been studied before \cite{demkowicz2015quantum,pinel2012ultimate,holland1993interferometric} and are therefore not of interest here. Other operators we use are the displacement operator $\hat{D}(\beta)$, the phase shift $e^{i\hat{n}\theta_p}$, and the identity operator $\hat{\mathbb{I}}$; the latter is included because we are promoting the easiest-to-implement schemes which would contain as many identities as possible. The final operator is constructed by performing a measurement and then inputting a new state, $|\psi_{new} \rangle \langle \psi_{meas}|$. This operator works by firstly implementing any one of the post-selection measurements described below, and then inputting one of the allowed input states $|\psi_0\rangle$ into this mode.

The final step of the state engineering scheme is to perform a post-selection measurement on one mode of the final state. If, for example, we wish to post-select onto the one photon state, we can perform a photon-number resolving detection (PNRD), and only keep runs which measure one photon. A measurement outcome of one photon therefore heralds the desired final state; later we discuss the consequences of this probabilistic heralding for the application of quantum metrology. The post-selection measurement corresponds to acting on the two-mode, pre-measurement state with $\langle 1|\otimes \hat{\mathbb{I}}$, followed by normalisation. We are then left with the single mode final state $|\psi_f\rangle$. Recent progress in PNRD has made detections of larger numbers of photons possible, and transition edge sensors can now resolve at least 4 photons to a reasonable efficiency \cite{humphreys2015tomography,gerrits2010generation}, whereas simpler detectors can reliably measure 1 or 2 photons \cite{achilles2004photon,fitch2003photon,zhou2015quantum}. The post-selection number measurements we include in the toolbox are therefore $\langle 1|$, $\langle 2|$, $\langle 3|$ and $\langle 4|$. Alternatively we can perform a quadrature measurement, $\langle x_{\lambda}|$, which is achieved with Homodyne detection, preceded by a phase shift which allows the quadrature phase to be controlled.
%The final measurement is the on-off detector which is significantly easier to implement than PNRD: the detectors clicks when it receive any number of photons and needs no number-resolving capabilities \cite{parigi2007probing,achilles2003fiber,morin2014non}.

%===========Fig:Random_general_scheme================
%====================================

%==========TABLE===========%

\begin{table}
\begin{center}
    \begin{tabular}{ c | c | c }
    Inputs, $|\psi_0\rangle$ & Operators & Post-selection measurements \\ \hline
    $|z\rangle$ & $\hat{U}_{T}$ & $\langle x_{\lambda}|$ \\
    $|\alpha\rangle$ & $\hat{D}(\beta)$ & $\langle 1|$ \\ 
    $|0\rangle$ & $e^{i\hat{n}\theta_p}$ & $\langle 2|$ \\
    $|1\rangle$ & $\hat{\mathbb{I}}$ & $\langle 3|$ \\
    $|2\rangle$ & $|\psi_{new} \rangle \langle \psi_{meas}|$ & $\langle 4|$ \\
    
    \end{tabular}  
\end{center}
\caption{This table summarises the quantum optics toolbox we use. Our algorithm selects elements from this toolbox and inserts them into the scheme in Fig.~\ref{fig:Random_general_scheme} in order to engineer non-classical states with the desired properties. See main text and Appendix A for details of the input states, operators and post-selection measurements.} 
\label{tab:toolbox}
\end{table}

%==========================%

We have described the elements of the quantum optics toolbox; we now introduce our algorithm - which uses the main principles of an evolutionary algorithm - for combining the various elements to engineer quantum states with the desired properties. Evolutionary algorithms \cite{eiben2003introduction,foster2001evolutionary,back1993overview,stepney2008searching} begin by creating an initial population of `chromosomes': each chromosome is a string which encodes a potential solution to the problem at hand. A fitness function, which returns a fitness value, is then used to identify the solutions which best solve the problem. The solutions with a large fitness value are retained and labeled the `parents', whereas the solutions with insufficient fitness values are discarded. The parents then breed the next generation of `offspring' which is achieved by `mutating' (making random changes to) the chromosomes of the parents. Often in evolutionary algorithms the chromosomes of two parents are combined, but we choose not to include this step in our algorithm. Our algorithm is therefore similar to a random search, but we retain the language of evolutionary algorithms for clarity of explanation.

After generating the offspring, the fitness function is applied to the offspring, and again the best solutions are selected from their fitness values. This identifies the next set of parents, and the process repeats until a collection of the `fittest' individuals remain. Evolutionary algorithms \cite{eiben2003introduction,foster2001evolutionary,back1993overview} have been successfully utilised to design antennas \cite{feichtner2012evolutionary,lohn2003evolutionary}, shape laser pulses \cite{baumert1997femtosecond}, and find new quantum algorithms \cite{stepney2008searching,spector1999finding}.

\begin{figure}%[h]
\centering
\includegraphics[scale=0.95]{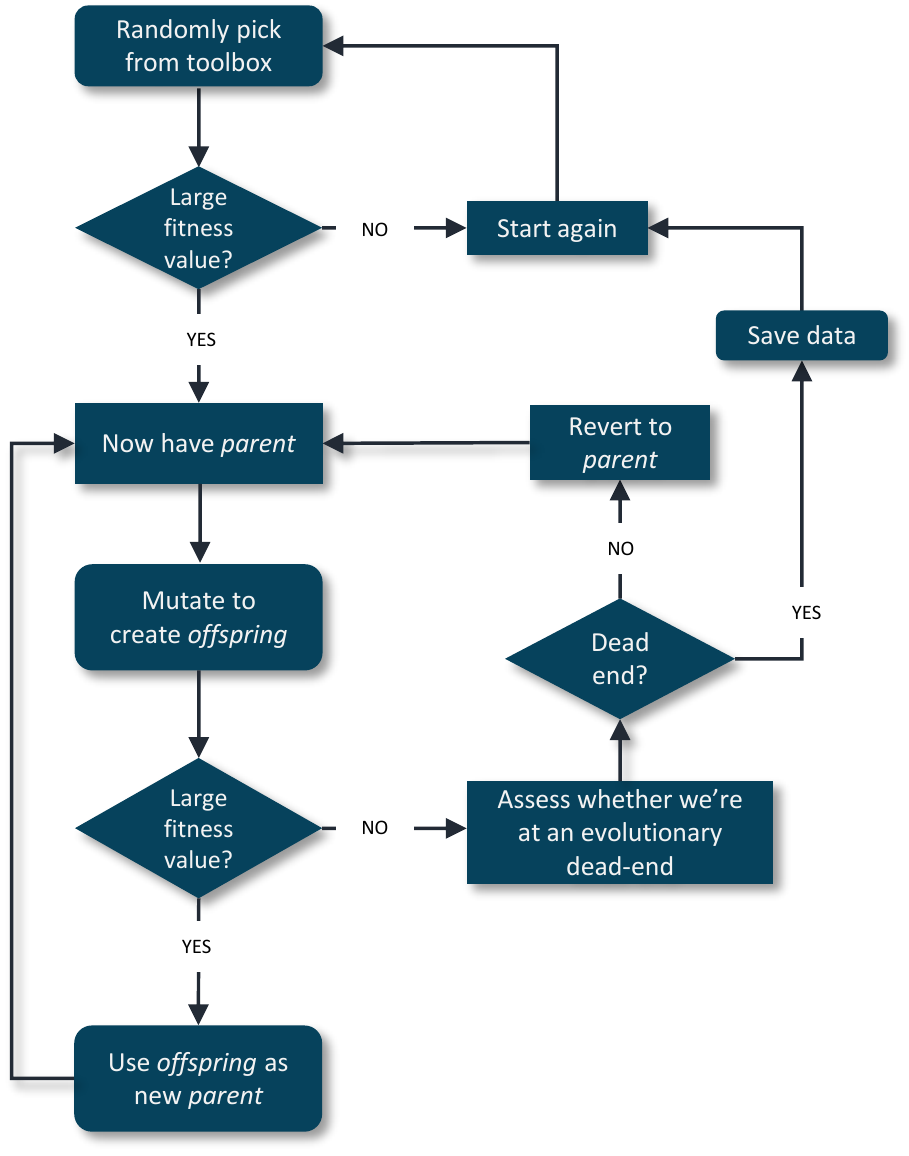}
\caption{Flow chart of \textit{Tachikoma}. Our algorithm \textit{Tachikoma} picks elements from the quantum optics toolbox in table \ref{tab:toolbox} to engineer quantum states, as described in the main text.  To assess the suitability of different states we use a fitness function: in this paper we use the quantum Fisher information as the fitness function in order to find states suitable for quantum metrology, but in principle any fitness function can be used which would allow us to find states for different applications.}
\label{fig:Flow_chart}
\end{figure}

A flow chart of our algorithm, which we name \emph{Tachikoma}, is given in Fig.~\ref{fig:Flow_chart}. Our initial population is created by randomly selecting inputs, operators and measurements from the quantum optics toolbox in table \ref{tab:toolbox}, and suitable elements of the initial population are selected according to their fitness value. The fitness function should therefore be chosen based on the properties the user desires for their quantum states. In this paper our fitness function is the quantum Fisher information (QFI), as a state with a large QFI can measure a phase shift to a high precision.

As shown in Fig.~\ref{fig:Flow_chart}, once a suitable parent is selected from the initial population we mutate it to create an offspring. If the fitness value of the offspring is greater than that of the parent, the offspring becomes the new parent for the next generation, and if not we revert back to the parent and repeat this process. If the offspring repeatedly fail to surpass their parent (an evolutionary dead-end) then we have likely reached a local maxima and so we store the details of the parent for further analysis and go back to the start of the algorithm. In this manner \emph{Tachikoma} mimics natural selection by progressively `evolving' quantum states which become increasingly suited for their purpose. \\

%===========Fig:Flow_chart================
%====================================

%================== Quantum Metrology ===================%

\section{Application to optical quantum metrology}

We now show how our algorithm \textit{Tachikoma} can be used to find states for optical quantum metrology. Specifically, we wish to find experimentally accessible states with low photon numbers. This is of relevance for measurements on fragile systems, with examples including spin ensembles \cite{wolfgramm2013entanglement}, biological systems \cite{carlton2010fast,taylor2013biological,crespi2012measuring,whittaker2015quantum} and atoms \cite{tey2008strong,eckert2008quantum}. In these examples, the number of photons we are able to interact with our system should be limited to prevent damage to the sample \cite{wolfgramm2013entanglement,taylor2015quantum}. The relevant resource here is therefore the number of photons in the probe state: to compare different states we look at their phase-measuring capabilities, given the same average photon number. Other methods of resource counting can be relevant, such as the total number of photons used to \textit{prepare} a quantum state, but we choose not to include these methods here.

\begin{figure}[t]
\centering
\includegraphics[scale=1.3]{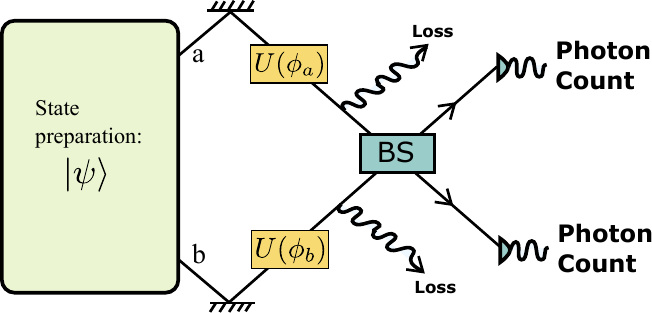}
\caption{This schematic shows how a phase shift can be measured in interferometry. A quantum state $\ket{\psi}$ is prepared and undergoes an unknown relative phase shift $\phi \equiv \phi_a - \phi_b$, which is applied with the operator $\hat{U}=\exp(i (\phi_a \hat{n}_a +\phi_b \hat{n}_b))$. For the states we consider the optimal measurement scheme is mixing the modes on a balanced (50:50) beam splitter (BS), followed by photon number counting. When photon losses are included these can be modeled by `fictitious' variable transmissivity beam splitters after the phase shift.}
\label{fig:Interferometer_general_input}
\end{figure}

We consider the standard optical phase estimation problem of measuring a phase difference $\phi$ between two optical modes containing unknown linear phase shifts, as shown in Fig.~\ref{fig:Interferometer_general_input}.  The fundamental limit to the precision with which a state $\rho$ can measure the phase $\phi$ is given by the quantum Cram\'er-Rao bound (CRB) \citep{braunstein1994statistical,braunstein1996generalized}:
\begin{eqnarray}
\delta \phi \ge \frac{1}{\sqrt{\mu F_Q(\rho)}}, \label{eq:CRB}
\end{eqnarray}
where $\mu$ is the number of independent repeats of the experiment and $F_Q(\rho)$ is the QFI of $\rho$. For pure states $|\psi\rangle$ the QFI is given by \cite{paris2009quantum,jarzyna2012quantum}
\begin{equation}
F_Q(\psi) = 4\text{Var}_{\psi}(\hat{G}) = 4 \left( \langle \hat{G}^2 \rangle -  \langle \hat{G} \rangle^2 \right),
\label{Eq:FQ1}\end{equation}
where $\hat{G}={1 \over 2} (\hat{n}_a - \hat{n}_b)$ is the generator of the phase shift $\phi \equiv \phi_a - \phi_b$, $\hat{n}_a$ ($\hat{n}_b$) is the photon number operator in mode $a$ ($b$), and the expectation values are taken with respect to the state $|\psi\rangle$. The QFI is used as our fitness function for the algorithm; we aim to maximise the QFI, which is equivalent to minimising the phase uncertainty.

We now introduce the shot noise limit (SNL) -- the best that can be done classically -- and the current state-of-the-art in optical quantum metrology. The SNL can be obtained by inputting a coherent state into a Mach-Zehnder interferometer. In the schematic in Fig.~\ref{fig:Interferometer_general_input} this amounts to using the input state $|\psi\rangle=\hat{U}_{T=50}|\alpha,0\rangle$ where $\hat{U}_{T=50}$ is a 50:50 beam splitter and $|\alpha\rangle$ is a coherent state, which are both defined in Appendix A. This state can measure at the SNL given by $\delta \phi = 1/\sqrt{\bar{n}}$, where $\bar{n}=\langle \hat{n}_a+\hat{n}_b \rangle$ is the total number of photons in the interferometer. A wide range of quantum states have been shown to surpass this, but to the best knowledge of the author the highest precision attainable by a practical state (a state that can be or has been made experimentally) is given by a pair of mode-separable squeezed vacuums (SV) $|\psi\rangle=|z,z\rangle$ where $|z\rangle$ is the squeezed vacuum state which again is defined in Appendix A \cite{demkowicz2015quantum,pinel2012ultimate}. This state can, among others, surpass Caves's squeezed state scheme \cite{caves1981quantum,pezze2008mach}, the NOON state \cite{lee2002quantum,afek2010high,jones2009magnetic}, and the recently produced squeezed cat states \cite{knott2015practical,ourjoumtsev2007generation,huang2015optical,etesse2015experimental} (see Appendix B for a discussion of this). The QFI of the SV and the SNL are shown in Fig.~\ref{fig:QFI_B1_6N1}; it is our goal in this paper to find practical states that can out-perform the SV.

%===========Fig:Interferometer_general_inputTMSC_with_loss================
%====================================

The algorithm we constructed excelled in this task and quickly found a number of states that have a significantly larger QFI than the SV. We highlight two states which stand out because they combine a large QFI with relative simplicity and experimental feasibility:
\begin{align}
|\psi_{T1}\rangle &= \mathcal{N}_{T1} \langle 2 | \hat{U}_{T=15} |z_1,z_2\rangle \\
|\psi_{T2}\rangle &= \mathcal{N}_{T2} \langle 3 | \hat{U}_{T=65} \hat{D}_a(\beta) |z_1,z_2\rangle,
\end{align}
where the subscript $a$ represents the first mode in Fig.~\ref{fig:Random_general_scheme} and $\mathcal{N}$ is the normalisation. We need a two-mode state to make a phase measurement using the scheme in Fig.~\ref{fig:Interferometer_general_input}, so we use $|\psi\rangle=|\psi_{Ti}\rangle \otimes |\psi_{Ti}\rangle$. The QFI of these states is plotted in Fig.~\ref{fig:QFI_B1_6N1} (specific values of the parameters, such as the phases and magnitudes of the squeezed states, are required for the results in Fig.~\ref{fig:QFI_B1_6N1}). We see that $|\psi_{T1}\rangle$ improves over the SV by a factor of 3 and, whilst being more difficult to implement, $|\psi_{T2}\rangle$ improves over the SV by a factor of 6.

All of the states found by \textit{Tachikoma} involve heralding, which is inherently probabilistic. The probabilities of success in producing $|\psi_{T1}\rangle$ and $|\psi_{T2}\rangle$ are approximately 0.1 and 0.09, respectively, with the exact probabilities depending on $\bar{n}$. Therefore, the experimental scheme for producing these states is only successful around $10\%$ of the time. A quantum metrology scheme using these states should produce them offline, and when the state is heralded by the appropriate measurement they can be inputted into the interferometer to measure the phase shift.

\begin{figure}%[h]
\centering
\includegraphics[scale=0.55]{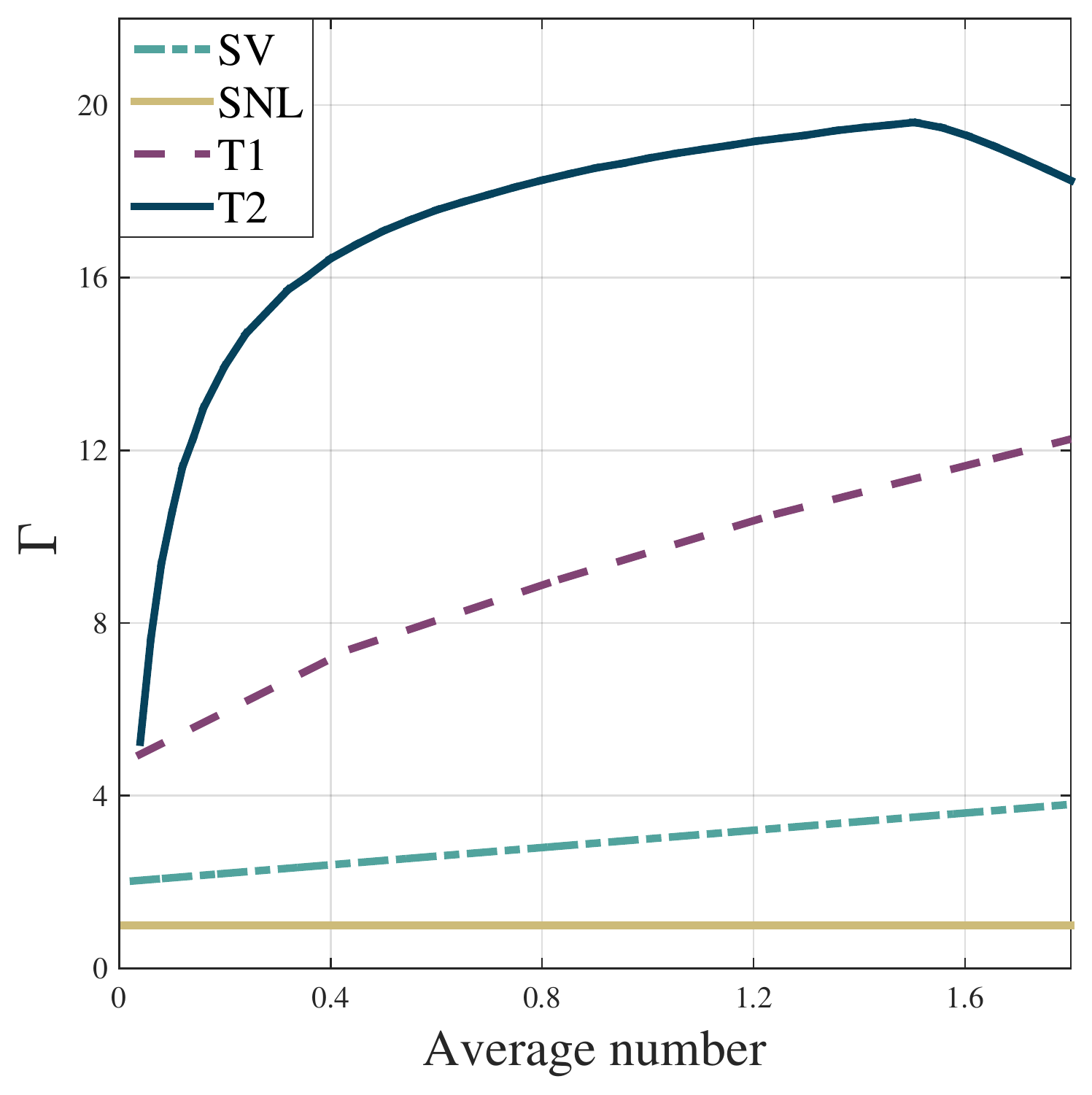}
\caption{Here we compare different states using $\Gamma = F_Q / \bar{n}$, which is the QFI ($F_Q$) scaled by the average number in the state ($\bar{n}$). Equation (\ref{eq:gary_gary}) shows that this measure can be used to directly compare the phase-measuring potential of different states. We see that $|\psi_{T1}\rangle$ and $|\psi_{T2}\rangle$ give significant improvements over the alternatives;  in particular $|\psi_{T2}\rangle$ can beat the SNL by more than $19$ times.}
\label{fig:gary_B1_6N1}
\end{figure}

While a large QFI does amount to a high precision it is not always the most revealing way to compare states. In this paper we are concerned with reducing the total number of photons that interact with the system being measured. We label the total number of photons (resources) $R$, which is given by $R=\bar{n}\mu$ where $\bar{n}$ is the average number of photons in the probe state and $\mu$ is the number of times this states is sent through the sample. We can then define a new measure which is more suited to the above restrictions: $\Gamma = F_Q / \bar{n}$. Then the CRB can be written as:
\begin{equation}
\label{eq:gary_gary}
\delta\phi \geq {1 \over \sqrt{R}} {1 \over \sqrt{\Gamma}} \propto  {1 \over \sqrt{\Gamma}}.
\end{equation}
We see that by fixing $R$ we can use $\Gamma$ to compare the performance of different states. In Fig.~\ref{fig:gary_B1_6N1} we plot $\Gamma$ against $\bar{n}$ for $|\psi_{T1}\rangle$ and $|\psi_{T2}\rangle$. We see that $|\psi_{T2}\rangle$ can improve over the SNL by more than $19$ times, which corresponds to more than a $4$-fold precision gain. Fig.~\ref{fig:gary_B1_6N1} illustrates that $|\psi_{T2}\rangle$ is optimal for $\bar{n} \approx 1.5$ (this isn't immediately clear by looking at the QFI). We can also see that for $|\psi_{T2}\rangle$ a large enhancement is possible for small states ($\bar{n} \approx 0.4$) which require smaller SV states to create and therefore are more practical.

%===========Fig:gary_B1_6N1================
%====================================

\emph{Tachikoma} also found the following states:
\begin{align}
|\psi_{T3}\rangle &= \mathcal{N}_{T3} \langle 3 | \hat{D}_a(\beta_1) \hat{D}_b(\beta_2) \hat{U}_{T=25} |0,z\rangle \\
|\psi_{T4}\rangle &= \mathcal{N}_{T5} \langle 4 | \hat{U}_{T=55} \hat{D}_a(\beta) \notag |z_1,z_2\rangle \notag\\
|\psi_{T5}\rangle &= \mathcal{N}_{T4} \langle x_{\lambda}=0 | \hat{U}_{T=95} |2,z\rangle \notag \\
|\psi_{T6} \rangle &= \mathcal{N}_{T6} \langle 1 | \hat{U}_{T=75} |\alpha,z\rangle. \notag
\end{align}
All these states improve over the SV by at least a factor of two. We will see below that $|\psi_{T3}\rangle$ is the most robust to photon losses, and without loss it has a QFI between $|\psi_{T1}\rangle$ and $|\psi_{T2}\rangle$. $|\psi_{T4}\rangle$ has a QFI even greater than $|\psi_{T2}\rangle$ but involves resolving $4$ photons, whereas states $|\psi_{T5}\rangle$ and $|\psi_{T6} \rangle$ are the easiest to implement.

The results of the algorithm are sometimes unexpected. For example, the operator $|\psi_{new} \rangle \langle \psi_{meas}|$ didn't give a significant enhancement and the simpler schemes were just as effective. States $|\psi_{T1-6}\rangle$ therefore only require linear operators and post-selective measurements to be made. Also, we experimented with between $m=2$ and $m=12$ operators $\hat{O}_i$ in the engineering scheme, but increasing the number of operators only made small improvements, and in $|\psi_{T1-6}\rangle$ only 1 or 2 operators are needed. \emph{Tachikoma} is efficient to run: after calibration it only took a few days of running on a single node of the HPC cluster at the University of Sussex to produce all the states presented in this paper. Longer runs just produced more copies of the same states. \\

%================== The Q Met Protocol ===================%

\section{The measurement scheme}

We wish to use our states to measure a phase shift using the setup in Fig.~\ref{fig:Interferometer_general_input}. We therefore require a pair of identical states to input into the two paths, $|\psi\rangle = |\psi_{Ti}\rangle \otimes |\psi_{Ti}\rangle$, where $|\psi_{Ti}\rangle$ is one of the states produced by \emph{Tachikoma}, or can be an SV for comparison (note that in this paper we take $\bar{n}$ to be the total number of photons in the state $|\psi_{Ti}\rangle \otimes |\psi_{Ti}\rangle$). We have quantified the performance of different states using the QFI and CRB, but it is important to now address the limitations of using the QFI as a figure of merit in quantum metrology \cite{hall2012universality,hall2012heisenberg,giovannetti2012sub}. In general, the precision as obtained by the QFI is achievable with an asymptotically large number of repeats, $\mu$. However, from a practical point of view it is clear that only some finite number of repeats $\mu$ will be possible. To factor this in we have performed a Bayesian simulation of the proposed experiment. Using the measurement scheme in Fig.~\ref{fig:Interferometer_general_input}, which involves mixing the modes on a balanced beam splitter followed by PNRD (see Appendix A for details of the PNRD), we have determined the phase shift, from a flat prior knowledge, using the Bayesian approach described in \cite{knott2015practical}. For all path-symmetric pure states (i.e. the states in this paper) the measurement scheme in Fig.~\ref{fig:Interferometer_general_input} is optimal and saturates the QFI \cite{hofmann2009all}. Indeed, our Bayesian simulation confirms that we come close to saturating the absolute bound given by the QFI for $\mu=O(10^2)$. In such regimes it is then clear that the states produced by \emph{Tachikoma} can significantly outperform the SNL and SV, in terms of absolute phase precision, when assuming the same average photon number. \\

%===========The effect of loss==============%

\section{The effects of loss}

We next investigate the effects of loss on the states produced by \emph{Tachikoma}. Loss can be modeled by adding `fictitious' beam splitters after the phase shift \cite{leonhardt1993realistic,demkowicz2009quantum}, as shown in Fig.~\ref{fig:Interferometer_general_input}, and results in a mixed state $\rho$ which typically has a reduced non-classical enhancement. The QFI for a general density matrix $\rho$ can be expressed as \cite{braunstein1994statistical,braunstein1996generalized,knott2015robust}
\begin{equation}
F_Q(\rho) = \sum_{i,j} \frac{2}{\lambda_i+\lambda_j}\left| \langle \lambda_i | \partial \rho(\phi) / \partial \phi| \lambda_j \rangle\right|^2,
\label{QFI}
\end{equation}
where $\lambda_i$ are the eigenvalues and $|\lambda_i \rangle$ a corresponding set of orthonormal eigenvectors of $\rho$.

%===========Fig:QFI_loss_J1_n0415_vs_SV_different_sizes================
\begin{figure}%[h]
\centering
\includegraphics[scale=0.54]{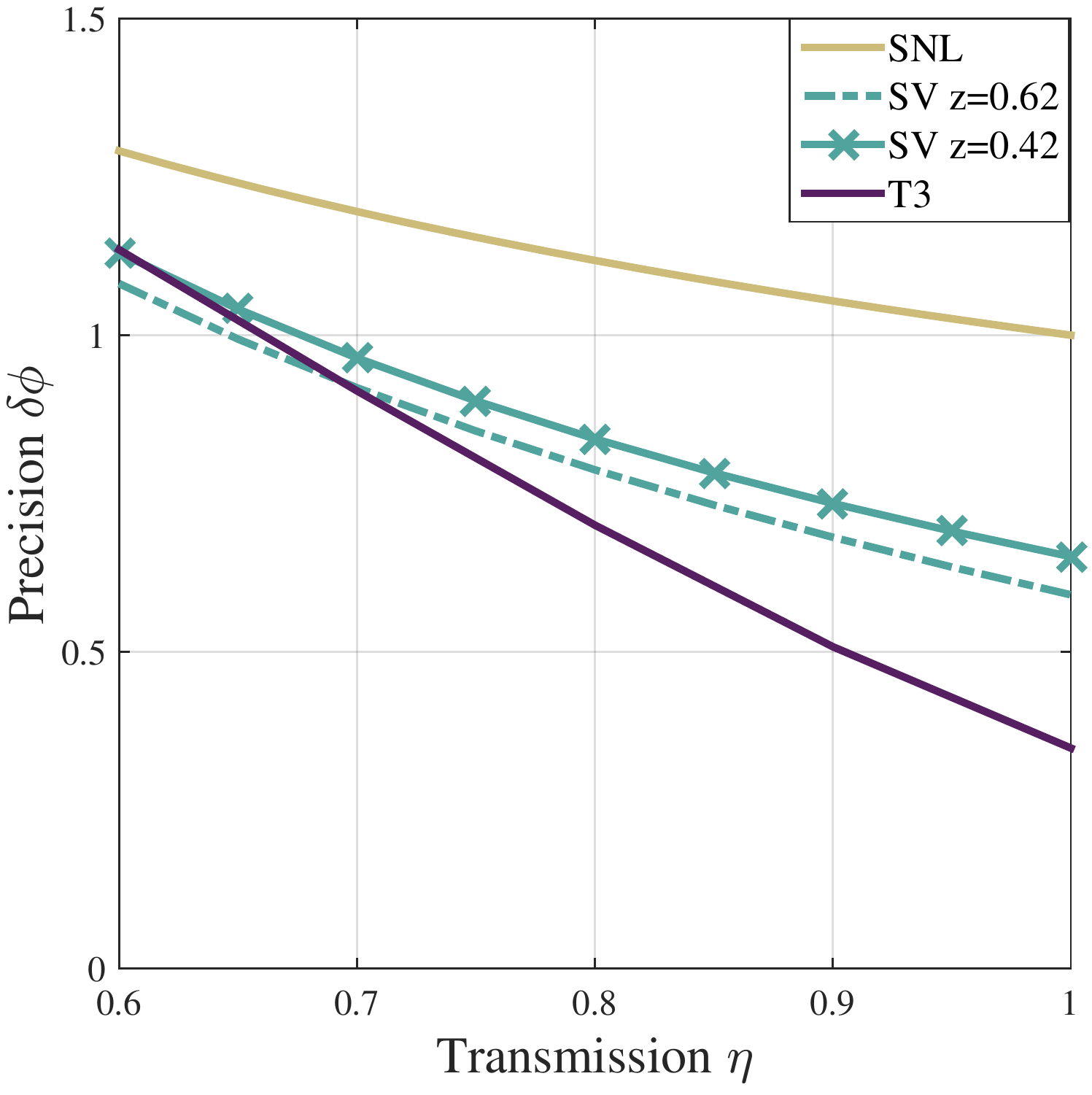}
\caption{The transmission probability through the interferometer, $\eta$, is plotted against the precision, $\delta\phi$ (scaled by $\mu$), for various states. The precision is found using equation (\ref{eq:CRB}). We see that the state $|\psi_{T3}\rangle$, found by our algorithm, shows significant improvements over the SNL and SV, even with moderate to high photon losses. We explain in the main text why we include two different-sized SV states.}
\label{fig:QFI_loss_J1_n0415_vs_SV_different_sizes}
\end{figure}
%====================================

The states $|\psi_{T1-6}\rangle$ have an intrinsic robustness to loss because they contain small numbers of photons and they are separable between the modes. As a result they all improve over the SV up to at least $15\%$ loss, which is the relevant loss rate for a number of experiments: losses as low as $10\%$ have already been achieved in table top interferometry experiments \cite{eberle2010quantum}, and near-future gravitational wave detectors are expected to have total losses of $9-17\%$ \cite{oelker2014squeezed}. The most robust state is $|\psi_{T3}\rangle$, which is shown in Fig.~\ref{fig:QFI_loss_J1_n0415_vs_SV_different_sizes}. We see that $|\psi_{T3}\rangle$ improves over the SV for losses up to $30\%$ and can beat the SNL with up to $50\%$ loss.

In Fig.~\ref{fig:QFI_loss_J1_n0415_vs_SV_different_sizes} we have included two different SV states because there are a number of valid comparisons that can be made. The first SV ($z=0.42$) has the same average number $\bar{n}$ as $|\psi_{T3}\rangle$. However, it can be disputed whether this is a fair comparison, because in order to create $|\psi_{T3}\rangle$ an SV is required with $z=0.62$, and we therefore also plot this state. We see in Fig.~\ref{fig:QFI_loss_J1_n0415_vs_SV_different_sizes} that $|\psi_{T3}\rangle$ provides a significant precision enhancement over both of these SV states.

\section{Discussion}

We have seen that our algorithm \textit{Tachikoma} has been constructed so that it can be easily edited to find quantum states for applications other than quantum metrology. The crucial change would be to the fitness function that we use to select successful states: here we use the QFI, but other measures can be used in order to make states suitable for quantum cryptography \cite{hillery2000quantum}, quantum computing \cite{lloyd1999quantum,kok2010introduction,munro2005efficient}, quantum teleportation \cite{pirandola2006quantum}, or boson sampling \cite{olson2015sampling}. We could also extend \textit{Tachikoma} to utilise quantum state engineering techniques that we have omitted in this paper, mainly due to practicality: we could create a $3$-mode entangled state before post-selection \cite{bimbard2010quantum}, include feed forwarding \cite{miwa2014exploring}, or look at cavity systems which allow for different operations to be performed \cite{vlastakis2013deterministically}.

In conclusion, we have created an algorithm that can be used to find optical quantum states with specific properties. In this paper we have focused on using the algorithm for quantum metrology, and we have found states that can surpass the best-known practical states by a factor of $\sqrt{6}$ in the precision, which amounts to over a 4-fold improvement over the classical shot noise limit. The states are experimentally accessible, robust to photon losses, and can be utilised for precise phase-measurements using a conceptually simple measurement scheme. We therefore expect that an experiment could confirm these results in the near future.

%=========Acknowledgments============
\section{Acknowledgements}
The author would like to thank Jacob Dunningham, Timothy Proctor, Adam Stokes, Robert Bennett and Konstantinos Meihanetzidis for helpful discussions. Nearing completion of this work we became aware of related techniques in \cite{krenn2015automated}. This work was funded by the UK EPSRC through the Quantum Technology Hub: Networked Quantum Information Technology (grant reference EP/M013243/1).
%===================================

%%====Appendix====%%

%\section{Methods}

%\begingroup
%\fontsize{8pt}{10pt}\selectfont
\section{Appendix A: Quantum optics toolbox details}
\label{sec:apx_toolbox}

\noindent\emph{Input states -} The squeezed vacuum is given by $|z\rangle = \hat{S}(z) |0\rangle$ where the squeezing operator is $\hat{S}(z)=\exp{ \left[ {1 \over 2} (z^* {\hat{a}}^{^2} - z {\hat{a}^{{\dagger}^2}}) \right] }$ and $z=r e^{i \theta_s}$ where $r$ is the (positive and real) amplitude, $\theta_s \in [0,2\pi]$ is the squeezing angle and $\hat{a}$ ($\hat{a}^{\dagger}$) is the annihilation (creation) operator. Squeezed states can be made up to $r \approx 1.4$, but this is extremely challenging experimentally so we set the limit to $r=1.3$ \cite{mehmet2011squeezed}. The coherent state is given by $|\alpha\rangle = \hat{D}(\alpha) |0\rangle$ where the displacement operator is $\hat{D}(\alpha) = \exp{ (\alpha \hat{a}^{\dagger} - \alpha^* \hat{a}) }$, $\alpha = |\alpha| e^{i \theta_c}$ where $|\alpha|$ is the  amplitude, and $\theta_c \in [0,2\pi]$ is the coherent state phase. The amplitude of the coherent state can be large in experiments, so instead it is limited by the numerical methods we use: we set the limit to $\alpha=4$. The final input state is the Fock state of which the simplest is the vacuum $|0\rangle$. Single photons, $|1\rangle$, can be emitted from a quantum dot \cite{muller2015coherent,claudon2010highly} or heralded \cite{morin2012high}. We also consider the two photon state, $|2\rangle$, which has been made in \cite{ourjoumtsev2006quantum,huang2015thesis}. Higher number Fock states can be made, e.g. by heralding, but we consider these states to be too difficult to produce reliably. \\

\noindent\emph{Operators -} The beam splitter is described by the unitary operator $\hat{U}_{T}=e^{-i\theta_b(e^{i\phi_b}\hat{a}^{\dagger}\hat{b}+e^{-i\phi_b}\hat{a}\hat{b}^{\dagger})}$, where $\hat{a}$ and $\hat{b}$ are annihilation operators for the two modes, and we choose the arbitrary phase to be $\phi_b=-\pi/2$. Here $T=100\cos^2{\theta_b}$ is the transmissivity of the beam splitter (in \%) and therefore for a 50:50 beam splitter $\theta_b=\pi/4$ giving $\hat{U}_{T=50}$. Next, the displacement operator, $\hat{D}(\beta)$ (defined above), is implemented by mixing the state with a large local oscillator at a highly transmissive beam splitter \cite{paris1996displacement} ($\beta$ has the same restrictions as $\alpha$). The phase operator is given by $e^{i\hat{n}\theta_p}$ where $\hat{n}=\hat{a}^{\dagger}\hat{a}$ and $\theta_p \in [0,2\pi]$. The identity operator is as expected. The final operator is constructed by performing a measurement and then inputting a new state and is given by $|\psi_{new} \rangle \langle \psi_{meas}|$ where $|\psi_{meas} \rangle$ is the post-selective measurement state, and $|\psi_{new}\rangle$ is the new state. \\

\noindent\emph{Measurements -} After we have applied a number of operators we perform a post-selective measurement on one mode of the final state. For example, if we wish to post-select onto the one photon state we can perform a number resolving detection (details below), and only keep runs which measure one photon. The measurement is given by a projection \cite{nielsen2010quantum}: to follow the single photon example we project with $|1\rangle \langle 1| \otimes \hat{\mathbb{I}}$. We are then left with a separable state $|1\rangle \otimes |\psi_f\rangle$, but we can ignore the measurement mode, and after normalisation we are left with the final one mode state: $|\psi_f\rangle$. This whole process can be more easily modeled by acting on the two-mode, pre-measurement state with $\langle 1| \otimes \hat{\mathbb{I}}$. In the main text we drop the identity and just write $\langle 1|$, and this measurement is always performed on the first mode of Fig.~\ref{fig:Random_general_scheme}.
%There is an exception when it comes to the on-off detector, which will be described below.

The quadrature measurement can be performed by Homodyne detection, preceded by a phase shift which allows the quadrature phase to be controlled. The eigenstates of the quadrature operator are given in \cite{barnett2002methods} by $|x_{\lambda}\rangle = \sum_{n=0}^{\infty} \langle n | x_{\lambda} \rangle |n\rangle$ where $\lambda$ is the quadrature angle, $x_{\lambda}$ is the quadrature eigenvalue, and the wave function is
\begin{align}
\langle x_{\lambda} | n \rangle = {1 \over \pi^{1/4}} {1 \over 2^{n/2}(n!)^{1/2}} e^{(-x_{\lambda}^2/2)} H_n(x_{\lambda}) e^{(-i n \lambda)}  \notag,
\end{align}
where $H_n(x_{\lambda})$ is the Hermite polynomial of order n. This quadrature post-selection is therefore given by $\langle x_{\lambda}|$, and has been implemented in \cite{etesse2015experimental,ourjoumtsev2007generation}. Note that we assume a perfect quadrature measurement here whereas in an experiment we would have to detect $x_{\lambda}$ in a certain small range.

Next we consider performing a photon-number resolving detection (PNRD). Recent progress in PNRD has made larger number detections possible and transition edge sensors can now resolve at least 4 photons to a reasonable efficiency \cite{humphreys2015tomography,gerrits2010generation}. Somewhat simpler detectors can measure 1 or 2 photons, for example by using time-multiplexing \cite{achilles2004photon,fitch2003photon} or a fan-out detector \cite{zhou2015quantum}. Our number post-selection number measurements are therefore $\langle 1|$, $\langle 2|$, $\langle 3|$ and $\langle 4|$.
%The final measurement is the on-off detector which is significantly easier to implement than PNRDs: they click when they receive any number of photons and need no number-resolving capabilities \cite{parigi2007probing,achilles2003fiber,morin2014non}. They are described by the projection $\sum_{m=1}^{\infty}|m\rangle\langle m| = \hat{\mathbb{I}} - |0\rangle \langle 0|$. Unlike the other post-selection measurements we use, the on-off detector creates a two mode state, and we can't just remove the second mode as in general the modes are entangled. We therefore have to treat this state differently than the others here and the two mode nature of the state after the on-off detector means than is it difficult to use this during the $|\psi_{new} \rangle \langle \psi_{meas}|$ operator, and so we omit it there.

\section{Appendix B: Beating the squeezed cat state}
\label{sec:apx_scs}

Here we have claimed that the states $|\psi_{T1-6}\rangle$ improve over the previously best known practical state, the SV, by up to a factor of $6$. We note here that recent work has shown that squeezed cat states (SCSs), given by $|\psi_{SCS}\rangle = \mathcal{N} S(z)(|\alpha\rangle + |-\alpha\rangle)$, can improve over the QFI of the SV by a factor of 3 \cite{knott2015practical}. Squeezed cat states have been made in \cite{ourjoumtsev2007generation,huang2015optical,etesse2015experimental}, so the reader may be lead to believe that our claim of beating the best known practical state by a factor of 6 is invalid. However, on closer investigation of \cite{ourjoumtsev2007generation,huang2015optical,etesse2015experimental} the squeezed cat states that have been made experimentally have a QFI even lower than the SNL. The reason for this is that to obtain a high QFI the parameters $z$ and $\alpha$ but be optimised over, but the protocols in \cite{ourjoumtsev2007generation,huang2015optical,etesse2015experimental} are unable to do this.

%\endgroup

%=============Bibliography============

\section{References}
%\nocite{apsrev41Control}
%\bibliographystyle{apsrev4-1}
%\bibliographystyle{apsrev}
%\small
\bibliography{MyLibrary_Thes_SES_RQM_Aug2015}

\end{document}